\documentclass[twoside,11pt]{Latex/Classes/PhDthesisPSnPDF}

\usepackage[T1]{fontenc}
\usepackage{array}
\usepackage{pdfpages}

\usepackage{multirow}
\usepackage{placeins}

\usepackage{xcolor}
\usepackage{framed}

\usepackage{graphicx}
\usepackage{url}

\usepackage{float}

\usepackage{graphicx}

\usepackage{mdframed}

\begin{document}

\thispagestyle{empty}

\begin{center}

Vrije Universiteit Amsterdam

\vspace{1mm}

\includegraphics[height=28mm]{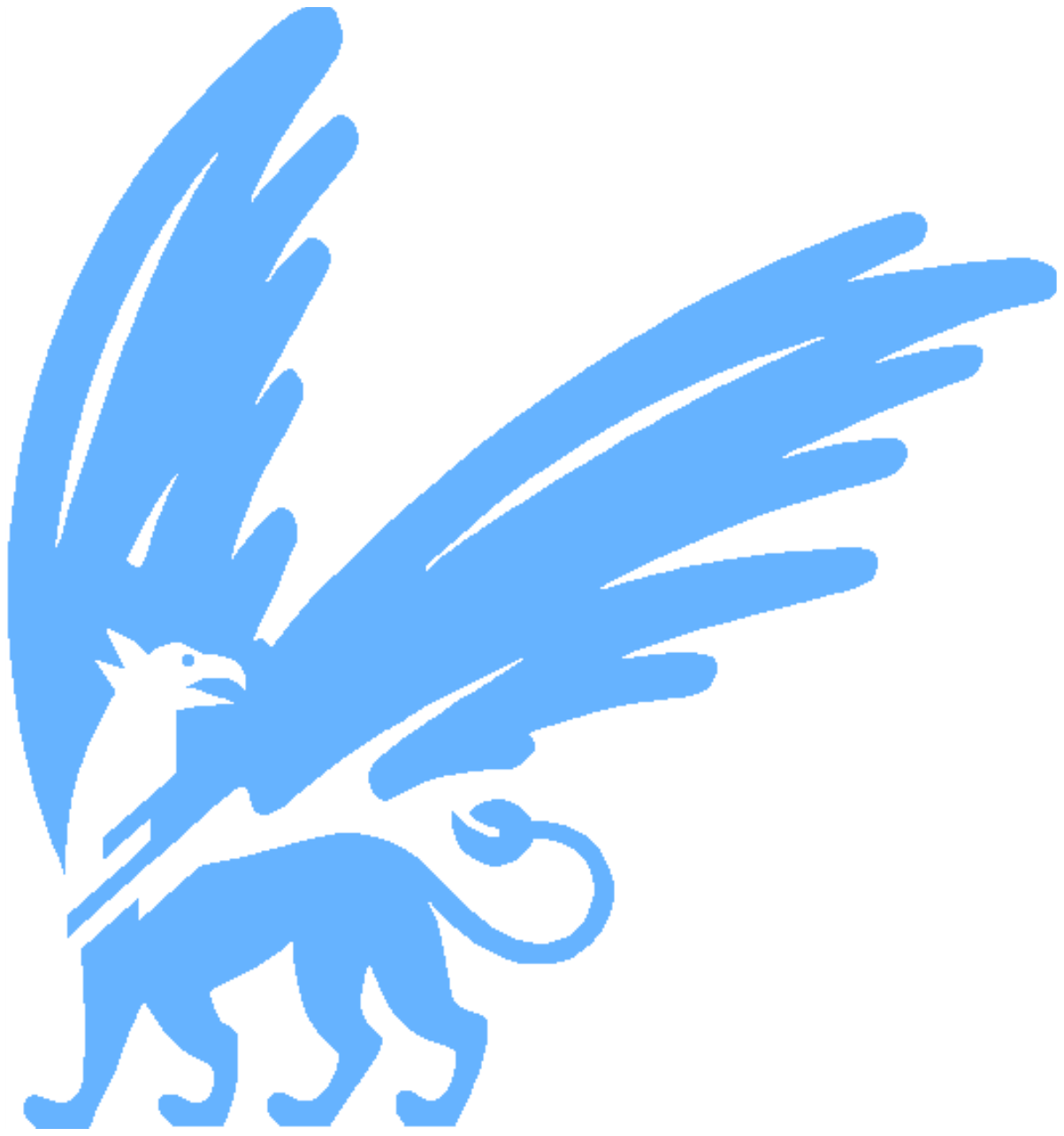}

\vspace{1.5cm}

{\Large Honours Programme, Project Report}

\vspace*{1.5cm}

\rule{.9\linewidth}{.6pt}\\[0.4cm]
{\huge \bfseries Failure Analysis of Big Cloud Service\par}
{\huge \bfseries Providers Prior to and During \par Covid-19 Period \par}\vspace{0.4cm}
\rule{.9\linewidth}{.6pt}\\[1.5cm]

\vspace*{2mm}

{\Large
\begin{tabular}{l}
{\bf Author:} ~~Muhammad Ahsan ~~~~ (2663138)
\end{tabular}
}

\vspace*{1.5cm}

\begin{tabular}{ll}
{\it 1st supervisor:}   & ~~prof. dr. ir. Alexandru Iosup \\
{\it daily supervisor:} & ~~Sacheendra Talluri, MSc\\
\end{tabular}

\vspace*{2cm}

\textit{A report submitted in fulfillment of the requirements for the Honours Programme, \\ which is an excellence annotation to the VU Bachelor of Science degree in\\ Computer Science/Artificial Intelligence/Information Sciences\\
version 1.0}

\vspace*{1cm}

\today\\[4cm] 

\end{center}

\newpage

       
\hbadness=10000
\hfuzz=50pt



\renewcommand\baselinestretch{1.2}
\baselineskip=18pt plus1pt




\newpage
%
%
%
%





\begin{abstracts}        

Cloud services are important for societal function such as healthcare, commerce, entertainment and education. 
Cloud can provide a variety of features such as increased collaboration and inexpensive computing. Failures are unavoidable in cloud services due to the large size and complexity, resulting in decreased reliability and efficiency. For example, due to bugs, many high-severity failures have been occurring in cloud infrastructure of popular providers, causing outages of several hours and the unrecoverable loss of user data \cite{Gunawi, Gunawi-failure-to-loss}. There are prior studies about cloud failure analyses are limited and use sources such as news articles. However, a detailed cloud failure focused study is required that provides analyses for cloud failure data gathered directly from the vendors.
Furthermore, the Covid-19 cloud failures should be studied as cloud services played a major role throughout the Covid-19 period, as individuals relied on cloud services for activities such as working from home. A program can be made for this task. As a result, we will be able to better understand and mitigate cloud failures to reduce the effect of cloud failures. 
\end{abstracts}



\frontmatter


\setcounter{secnumdepth}{3} 
\setcounter{tocdepth}{3}    
\tableofcontents            


\mainmatter

\renewcommand{\chaptername}{} 




\chapter{Introduction}

\ifpdf
    \graphicspath{{1_introduction/figures/PNG/}{1_introduction/figures/PDF/}{1_introduction/figures/}}
\else
    \graphicspath{{1_introduction/figures/EPS/}{1_introduction/figures/}}
\fi



Utilizing the cloud is a trend that continues to grow as it is used by an increasing number of individuals and businesses \cite{CloudAndEconomyGrowth, CloudRevenueGrowthNews}. Users may utilize cloud services when and where they need them and only pay for the resources they use. It provides massive computational capacity, scalability on demand, and utility-like availability at a reasonable cost. Cloud Computing enables efficiently on-demand network access to a shared pool of configurable computing resources that can be rapidly provisioned and released with minimal management effort or service provider interaction. The cloud is ubiquitous. However, cloud failures occur frequently and affect users of the cloud services. These failures can lead to great loss as people’s daily activities such as many peoples income depends on the cloud services to work.

\section{Context} 

A failure can be described as an unintended behavior of a service.
The term ``cloud failure'' refers to when a cloud service is unavailable for use for an extended period of time. Unavailability can also relate to a service's inadequacy in terms of performance, as measured by the agreed-upon SLA metrics. An SLA is an explicit or implicit contract between a cloud provider and customer. It governs the obligations and responsibilities between both parties regarding the provided service. In case the cloud provider is unable to meet the terms of the agreed SLAs, the consequence is usually a financial penalty or rebate, but it can take other forms as well. For example, if a data centre was only partially impacted by a failure, the vendor may be forced to do the appropriate repair and restoration actions. The end user may experience downtime until the service is fully restored in accordance with the agreed-upon SLA criteria.
Despite all the potential and development cloud computing has undergone
over the years, failures continue to occur. Cloud failures are of concern as they can result in reduction of quality of service, availability, reliability and energy waste \cite{energy-waste} that can ultimately lead to economic loss for both cloud users and providers. By studying the failures, we can protect users from
being affected by these failures.

There are numerous cloud providers. Cloud failures are common among all
providers. Big cloud companies represent a large population of cloud
users. Therefore, it is more beneficial to study failures in big cloud
providers. This is the aim of the current study. Many big cloud vendors
provide cloud failure logs that can be used for the study. Cloud
failures can be analyzed over time to gain better understanding.
Furthermore it is interesting to compare failures among different
vendors. Cloud failure analysis is done with the aid of a program.

There are studies that describe different aspects of cloud. However
there is limited in-depth research on the topic of cloud failures. This
paper will provide analysis about cloud failures in three popular big
cloud providers Amazon Web Service (AWS), Microsoft Azure and Google
Cloud Platform (GCP), and comparison of failures during the period
2018 till 2020 (June).

\section{Problem Statement}

As the dependency on cloud computing increases, society demands high availability, an ideal 24/7 service uptime, if possible. The cloud runs many daily life activities including banking, healthcare, governance, transportation, e-commerce, entertainment, etc. Cloud services played an important part during Covid-19. A large population of the world trusts cloud services to run their daily lives, for example, employees working remotely from homes, online education for students and online businesses. However, many cloud services continue to fail. These failures can lead to great loss as people's daily activities, especially as many people's income depends on the cloud services to work.

Is it correct to depend on the cloud to run daily lives? Why does the cloud keep failing? Which services are failing? Where, when and how are the failures occurring? How do the big cloud providers respond to failures? What is the pattern of the failures in different periods?  During difficult times like the Covid-19 period, can the cloud fulfil the needs of people. Many similar important questions can be explained by analysing the failures on cloud.

\newpage

\section{Research Questions}

\noindent \textbf{[RQ1] How to process raw cloud failure data to carry out analysis using the data? } 

The amount of data is substantial, since most files include approximately 1000 instances. Any manual approach is likely to make mistakes and is inefficient for solving the current problem because data is raw and would require customize-able processing. Therefore an efficient method to solve the issue of preparing data is to use an automated method which is reproducible. One of the possibilities is to use a program that does the task. However, no known program exists that can prepare the given sets of data.\newline
    
\noindent \textbf{[RQ2] How to statistically characterize and visualize cloud failure data to understand the frequency of failures? }

After approaching research question 1, we would have a software that outputs the final data sets. At this point some approach is required that can extract information and present the data. A possibility can be to provide the processed data as an input to some available data analysis software. A software like Microsoft Excel only provides basic ways to analyze and do not meet the requirement of the current study. For example, if the study requires a specific graph with specific features, the data analysis software may not have those. Additionally, there are three different data sets, doing manual analysis using some software for every data file is not efficient. In a nutshell, general data analysis softwares provide limited options.\newline

\noindent \textbf{[RQ3] How did the pattern of failures change when comparing the Covid-19 period to the prior years?}

Information such as peak cloud failures and their relation with other features such as time and location need to be extracted to gain valuable insight in cloud failures. Additionally, People relied on cloud services during Covid-19 example doing work from home. The change in usage can help us view failure patterns that differ from normal days. Therefore comparison of Covid-19 period cloud failures with the period before can be used to better understand cloud failure.

\newpage

\section{Approach}

\textbf{RQ1.} 
The question under consideration can be solved by creating a program that takes the raw data file as the input, prepares it such as  cleaning, filtering, organising etc. The program outputs a file with ready to use data. There are a variety of programming languages available for making the software. A preferable language can be python because there are many data-centric Python packages, for example Pandas and NumPy, which make the process of data processing and data analysis a lot quick and convenient.

\noindent \textbf{RQ2.} 
The current research requires an analysis approach that is efficient, reusable and provides a wide range of customizing options. The given requirements can be fulfilled by coding a software that does the job. This software can be an extension of the program made in research question 1. In this way, the data can directly be available for analyses. Moreover, as suggested in the approach to research question 1, python can be used as it provides a variety of options example for plotting matplotlib and seaborn plot can be used. 

\noindent \textbf{RQ3.} 
The first part of the question can be approached by understanding the general structure of the data files such that considering the data fields(columns of data set), it is possible to predict some possible analyses. The yearly data can be grouped by separate vendors. This can provide answers to questions like which vendor has the most failures. The failure event start time and date of failures can be used to analyze months that had the most failures. Using event start time with event end time we can find out the events that lasted the longest. This can further be used to analyse the performance of service providers. The location data can tell which location had the most failures. Similarly the service name column can be analyzed to get services that failed the most. A possible plot can be a plot to show failure count for each month with respect to separate cloud providers. 

The second part of the question can be answered such that, the Covid-19 reduced the physical interaction among people and increased interaction through cloud services, for example, online education and business. This rapid increase in demand increased the burden on cloud services. Considering the fact that Covid-19 was not expected, the cloud was not optimised for this, at least not initially. Therefore it can be predicted that more failures should have occurred during Covid-19 period. As a result, it is also interesting to compare and provide analyses for the year 2020 (Covid-19 period) and the previous years.

The study results in the following contributions:

\begin{enumerate}

\item[\textbf{[C1]}] A tool to process raw cloud failure data and extract useful information. The first research question build the software to clean raw cloud data. Then research question 2 extends the software from RQ1 adding visuals and statistic results. \newline
    
\item[\textbf{[C2]}] Analyses of cloud failures prior and during Covid-19 period. Additionally, cloud failure analysis can be used to reduce cloud failures and for further research.\newline

\end{enumerate}


\sloppy

{\let\cleardoublepage\relax \chapter{Processing Cloud Failure Data} \label{cleaning_data}


\ifpdf
    \graphicspath{{2_background/figures/PNG/}{2_background/figures/PDF/}{2_background/figures/}}
\else
    \graphicspath{{2_background/figures/EPS/}{2_background/figures/}}
\fi

%

This chapter describes the characteristics of the cloud failure data. The operations performed to process the data such as cleaning and organising data are also described in this chapter. Therefore, the first research question is answered in the chapter. The operations mentioned here are in order. If a process is mentioned before another process. It indicates that the process was done before the next process unless specifically mentioned in the description. Therefore the understanding of the next process should be gained assuming the processes mentioned before is already completed. The cloud failure data used by the study is obtained from the official
website of the big cloud providers \cite{Azuredata,GCPdata,AWSdata}. The study
covers the period 2018 till 2020 (June).

\section{Data Description} \label{data-description}

The failure information
was initially put in rows and gathered in three separate files for
each year named `provider failures 2018', `provider failures 2019' and
`provider failures 2020'. All three files had 12 columns 
that had information about service id, service name, location,
status, event start time, event end time, event duration hours,
first notification, last notification, description, vendor, monitor
and orgtype. Most column headers are self-explanatory. The event start
time, event end time, first notification and last notification column
has information about the date and time. This is originally represented in the
form of a Unix timestamp. The orgtype column means origin type. The
origin type is `cloud' for all rows as data is about cloud services. The
study covers three big cloud services so the vendor column has three
possibilities that are; AWS, Azure and GCP. The vendor and monitor columns contain
the same information. For consistency, the vendor column was used when
needed during the study. The program made for cloud analysis was used to extract information about the data. This
was done by choosing the column and outputting unique values and their
respective counts. Description of columns are shown in Table \ref{tab:col-description}.\newline

\begin{table}[]
\centering
\resizebox{\textwidth}{!}{%
\begin{tabular}{ll}
\hline
\textbf{Column name}          & \textbf{Description}                                             \\ \hline
service id           & The id of the service                                   \\
service name         & Name of service                                         \\
location             & Region of failure                                       \\
status               & Indicators such as 1,0. No meaningful interpretation    \\
event start time     & Start time of failure                                   \\
event end time       & End time of failure                                     \\
event duration hours & Derived feature representing duration of cloud failure  \\
first notification   & First notification of the failure issued by the vendor. \\
last notification    & Last notification of the failure issued by the vendor.  \\
description          & Details of cloud failure.                               \\
vendor               & The cloud provider AWS or GCP or Azure.                 \\
monitor              & Same as vendor.                                         \\
org type             & For all instances the origin is cloud.                  \\ \hline
\end{tabular}%
}
\caption{Data columns description.}
\label{tab:col-description}
\end{table}

The file `provider failures 2018' had 965 rows, file
`provider failures 2019' had 1024 rows and `provider failures 2020' had
639 rows. The rows of 2020 file are less compared to other
years files because 2020 covers cloud failures until June while other
files have full-year cloud failure data. The values mentioned are
of the raw data on which no operation had been performed. Each file
contains information about three big cloud vendors AWS, Azure and GCP.
In this initial data set rows were disorganised. The original failures count with respect to
each cloud provider is given in the first row of Table \ref{tab:dup-removal}. The data set contains
rows with missing data. There are three vendors and they do not provide information required to fill all 12
columns, see Section \ref{unknown-cells}. These cases are handled
carefully at the right time in the right way, see Section \ref{cleaning_data}.

Many operations are the same for all
three files because of the similar structure. When the tool is run, input is given to the program which specifies the file to process. Input 1 to process file
`provider failures 2020', 2 for file `provider failures 2019' and 3
for file `provider failures 2018'. When files are processed the program reads/stores the processed versions files. Reading all three files can enable comparing data such as combined plots. The
program prints the information to the screen. The
processed results are verified by automated tests built in the tool and manual checks. Manual
checks include observing different parts of the data before and after the processing. Multiple people are involved so error is minimized.
Depending on the tests, the program is updated if required. The study includes 2020 data till June. When 2020 is mentioned in this text, it means till June 2020.

\section{Removing Duplicate Data}
The initial data set had duplicate rows, described in Section \ref{data-description}. Removing duplicates is the first step to
prepare data. Duplicate rows are when a failure event is reported more
than once in the file. Duplicate rows have the same service\_id,
service\_name, location, event\_start\_time and vendor.

Description is not included in the criteria for duplicate rows.
Duplicate rows will have the same description because the event is the
same. However, the string length used for checking duplicates can be different. For example, two duplicate rows, the first row the description starts with ``The''
but the second row starts without the ``The'', the remaining description
is the same for both rows, will have different string length.
This example was observed during the study. The tool outputs the
duplicate rows which are used studied to verify the process
of duplicate removal. Similarly we ignore event end time when
removing duplicates. There is a special case in which the rows are
duplicate and have the same description length but have different end
times. The quantity of such rows is  few. In this case we keep
the row with higher end time. Comparing descriptions and event end times
of two rows would not be beneficial as seen in the example. Instead of
using description we use other fields (mentioned above) to check for
duplicates. The tool does the following to remove duplicates: starting at the first row it selects that row and compares with the
rows below the selected row in the file. If a duplicate row is found, it
is removed from the file. In normal cases, when two rows are the
same, the selected row is kept and the other rows are deleted. The Table \ref{tab:dup-removal} shows the result of removing duplicates. The plus 1 in Table means that 1 row had no vendor
information. In both 2019 and 2018 files, this row gets removed at a
later step, see Section \ref{subsec:incorrect-dates}.

\begin{table}[]
\centering
\resizebox{\textwidth}{!}{%
\begin{tabular}{l c c c c|c c c c|c c c c}
\hline
\begin{tabular}[c]{@{}l@{}}\bf{Number of} \\ \bf{Rows}\end{tabular} 
                                                                           & \multicolumn{4}{c|}{\bf{2018}}                                            & \multicolumn{4}{c|}{\bf{2019}}                                             & \multicolumn{4}{c}{\bf{2020}}                                            \\ 
                                                                           & \bf{AWS} & \bf{Azure} & \bf{GCP} & \bf{Total}                                             & \bf{AWS} & \bf{Azure} & \bf{GCP} & \bf{Total}                                               & \bf{AWS} & \bf{Azure} & \bf{GCP} & \bf{Total}                                             \\ \hline 
\bf{Initial}                                                                    & 325 & 284   & 355 & \begin{tabular}[c]{@{}c@{}}964\\ +1\end{tabular} & 273 & 222   & 528 & \begin{tabular}[c]{@{}c@{}}1024\\ +1\end{tabular} & 271 & 51    & 316 & \begin{tabular}[c]{@{}c@{}}638\\ +1\end{tabular} \\[3ex]
\bf{Duplicate}                                                                  & 4   & 145   & 142 & 291                                              & 7   & 110   & 261 & 378                                               & 1   & 15    & 114 & 130                                              \\[3ex]
\begin{tabular}[c]{@{}l@{}}\bf{After Removing} \\ \bf{Duplicates}\end{tabular}       & 321 & 139   & 213 & \begin{tabular}[c]{@{}c@{}}673\\ +1\end{tabular} & 266 & 112   & 267 & \begin{tabular}[c]{@{}c@{}}645\\ +1\end{tabular}  & 270 & 36    & 202 & \begin{tabular}[c]{@{}c@{}}507\\ +1\end{tabular} \\ \hline
\end{tabular}%
}
\caption{Results of duplicate removal.}
\label{tab:dup-removal}
\end{table}

\section{Handling Missing Data} \label{unknown-cells}
The unknown cells include the cells that are empty cells or cells with ‘-1’ or ‘0’. The reason is that not all vendors provide all information as addressed by the current study, see Section \ref{data-description}. Furthermore, for consistency we aim to replace the multiple representations and use ‘Unknown’ for string based unknown cells and 0 for integral based unknown cells. The unknown cells are GCP location cells in 2020 and 2019 data files, see Section \ref{subsec:incorrect-dates}. Another case is Azure does not have service id in any row in any year. Azure also does not provide information about the first notification and last notification. These are also the limitations of the data and the study. All other rows are filled with appropriate information. Table \ref{tab:missing-and-old-dated} gives the frequency of the unknown cells as per vendor and total.

\section{Incorrect Dates in Data} \label{subsec:incorrect-dates}
The initial data had rows with incorrect date and time. The rows had ``-1''
as event start time and event end time. ``-1'' indicates that event time
was not known. Due to lack of information, the rows with ``-1'' were
removed. There were also rows that had event time of the previous year,
for example, the file of 2020 contained  
failure events of 2019 and 2018. The old dated events were already
present in the correct year files and were repeated in the new year
file. The old dated rows were not required in the new year file so they
were removed from the file. The results are shown in table below.

\begin{table}[H]
\centering
\begin{tabular}{l c|l|l|l|c|l|l|l|c|l|l|l}
\hline
\begin{tabular}[c]{@{}l@{}}\bf{Number of} \bf{Rows}\end{tabular}      
                                                                               & \multicolumn{4}{c}{\bf{2018}} & \multicolumn{4}{c}{\bf{2019}} & \multicolumn{4}{c}{\bf{2020}} \\ \hline \\[0.1ex] 
\begin{tabular}[c]{@{}l@{}}\bf{After Removing}\\ \bf{Duplicates}\end{tabular}            & \multicolumn{4}{c}{674}  & \multicolumn{4}{c}{646}  & \multicolumn{4}{c}{509}   
\\[3ex]\begin{tabular}[c]{@{}l@{}}\bf{Incorrect dated}\\ \bf{Rows}\end{tabular}                 & \multicolumn{4}{c}{275}  & \multicolumn{4}{c}{271}  & \multicolumn{4}{c}{386}  
\\[3ex]\begin{tabular}[c]{@{}l@{}}\bf{After removing} \\ \bf{incorrect dated rows}\end{tabular} & \multicolumn{4}{c}{399}  & \multicolumn{4}{c}{375}  & \multicolumn{4}{c}{123}  \\ \hline
\end{tabular}
\caption{Missing dates and old dated rows.} 
\label{tab:missing-and-old-dated}
\end{table}

\section{Special cases in Data}
The initial files
also had other special cases related to event start, end date and time.
A few rows had the same event start and end time. Same event start and end
time is not possible so these rows are considered incorrect and removed.
Another case is that some rows had event end time before the event start
time. These rows were studied manually and checked with the time in the
description. A common thing that was noticed in rows having end time
before start time was that the start time was not converted
to a 24-hour format. We converted these to 24-hour time and the times
matched the timing mentioned in the description. In general these rare
cases were present in only a few rows. In total less than 15 rows per
file had end time before start time. In this case rows were kept in the
files fixed by converting to 24-hour format. Another case was that there
were few rows that had corrupted description. These rows count at most
5. Corrupted description means the rows description had symbols not
understandable English. These rows were removed. After taking care of
the special cases, the failures count for each files is shown in
Table \ref{tab:special-rows}.

\begin{table}[H]
\centering

\begin{tabular}{l c l|l|l|c|l|l|l|c|l|l|l}
\hline
\begin{tabular}[c]{@{}l@{}}\textbf{Number of} \textbf{Rows}\end{tabular}
                                                                               & \multicolumn{4}{c}{\textbf{2018}} & \multicolumn{4}{c}{\textbf{2019}} & \multicolumn{4}{c}{\textbf{2020}} \\ \hline \\[0.1ex] 
\begin{tabular}[c]{@{}l@{}}\textbf{After removing} \\ \textbf{incorrect dated rows}\end{tabular} & \multicolumn{4}{c}{399}  & \multicolumn{4}{c}{375}  & \multicolumn{4}{c}{123}  \\[3ex]
\textbf{Special cases rows}                                                             & \multicolumn{4}{c}{13}   & \multicolumn{4}{c}{18}   & \multicolumn{4}{c}{9}     \\[3ex] 
\textbf{Remaining rows}                                                                 & \multicolumn{4}{c}{386}  & \multicolumn{4}{c}{357}  & \multicolumn{4}{c}{114}  \\ \hline
\end{tabular}%

\caption{Result of fixing incorrect date and special cases.}
\label{tab:special-rows}
\end{table}

\section{Normalizing text}
After cleaning the data and sorting rows, the data was more readable. The quantity was reduced compared to initial files. At this stage data files were manually observed. It was noticed that in the column of location and service name, some rows needed to be fixed. Two rows have location ‘East US’ but the second row uses lowercase ‘east us’. It is likely that when performing comparisons between two cells, this can give incorrect results. Another scenario is that ‘Networking’ and ‘Network’ represent the same service name. Similarly ‘Southeast Asia’ and ‘South East Asia’ represent the same location. When two cells intend to deliver the same information, we replace them with a single representation, for example we replace  ‘Southeast Asia’ and any other representation of southeast asia with ‘South East Asia’. These measures are helpful in reading and sorting data. This results in single representation and provides the correct result when filtering, example when outputting unique values. This operation is done before removing duplicate rows, this improves the result of the duplicate removal process. The failure count presented in the tables such as Table \ref{tab:dup-removal} is result obtained after doing this process.

\begin{table}[h]
\centering
\resizebox{\textwidth}{!}{%
\begin{tabular}{l c c c c|c c c c|c c c c}
\hline
\begin{tabular}[c]{@{}l@{}}\bf{Number of} \\ \bf{Rows}\end{tabular}
                                                                           & \multicolumn{4}{c}{\bf{2018}}                                            & \multicolumn{4}{c}{\bf{2019}}                                             & \multicolumn{4}{c}{\bf{2020}}                                            \\ 
                                                                           & \bf{AWS} & \bf{Azure} & \bf{GCP} & \bf{Total}                                            & \bf{AWS} & \bf{Azure} & \bf{GCP} & \bf{Total}   
                                                                           & \bf{AWS} & \bf{Azure} & \bf{GCP} & \bf{Total}   
                                                                           \\ \hline
\bf{Initial}                                                                    & 325 & 284   & 355 & \begin{tabular}[c]{@{}c@{}}964\\+1\end{tabular} & 273 & 222   & 528 & \begin{tabular}[c]{@{}c@{}}1024\\+1\end{tabular} & 271 & 51    & 316 & \begin{tabular}[c]{@{}c@{}}638\\+1\end{tabular} \\[3ex]
\begin{tabular}[c]{@{}l@{}}\bf{Total} \\ \bf{Removed}\end{tabular}                   & 188 & 165   & 255 & \begin{tabular}[c]{@{}c@{}}578\\+1\end{tabular} & 155 & 144   & 367 & \begin{tabular}[c]{@{}c@{}}667\\+1\end{tabular}  & 233 & 34    & 257 & \begin{tabular}[c]{@{}c@{}}524\\+1\end{tabular}  \\[3ex] 
\begin{tabular}[c]{@{}l@{}}\bf{Remaining}\\ \bf{after} \\ \bf{cleaning}\end{tabular}     & 137 & 119   & 130 & 386                                              & 118 & 78    & 161 & 357                                               & 38  & 17    & 59  & 114                                              \\ \hline
\end{tabular}%
}
\caption{Frequencies of data removed during cleaning process.}
\label{tab:my-table}
\end{table}

\section{Organising Data}

The data rows in the initial files were disorganized. Generally, there were many sorting options. In particular, the order location, vendor, service id, event start time, and then service name was found to be the optimal sorting order. Rows are sorted by location first, then vendor, and so on. The information was arranged in ascending order. The sorting operation was very useful in certain analyses and also in outputting data.

The data has been cleaned and organized into groups. We have the data as separate year data files initially. We store data after it has been cleaned with respect to three different vendors for each year. This can be useful for analyses and in other operations, such as graph creation.

} 




{\let\cleardoublepage\relax \chapter{Results and Analysis}


\ifpdf
    \graphicspath{{4_implementation/figures/PNG/}{4_implementation/figures/PDF/}{4_implementation/figures/}}
\else
    \graphicspath{{4_implementation/figures/EPS/}{4_implementation/figures/}}
\fi

%

This chapter answers the second and third research question. To answer the second research question the data is visualized using plots. Furthermore, the plots are analyzed to extract patterns. A detailed comparison is laid down between the period before and during Covid-19 for each plot. This answers the third research question.

\section{Monthly Cloud Failures}

\begin{figure}[]
    \centering
    \includegraphics[width=\textwidth,keepaspectratio]{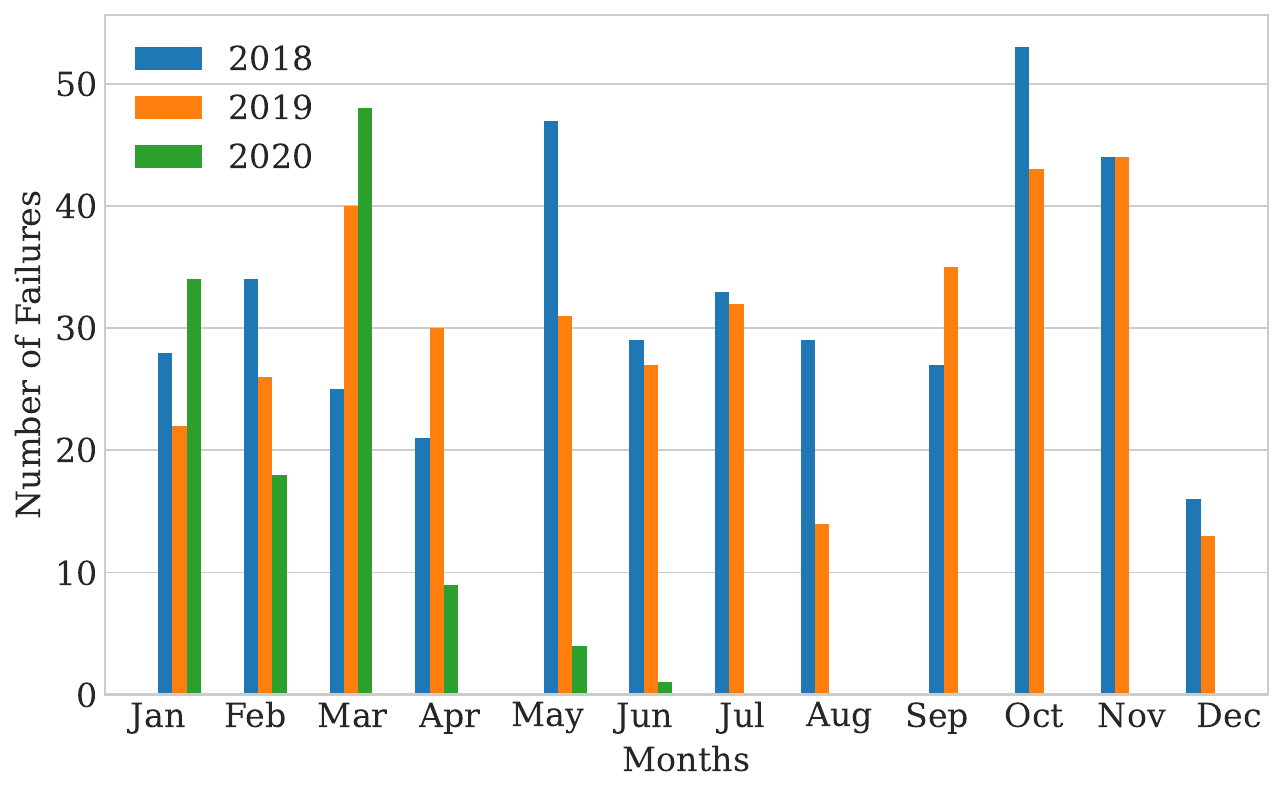}
    \caption{Plot of failure frequency in each month.}
    \label{fig:monthly-plot}
\end{figure}

The Figure \ref{fig:monthly-plot} shows the total number of monthly cloud failures in 2019 is less than in 2018 except for March, April, and September (2020 not included). That is, in 75 percent of cases, the cloud failures have decreased. In both 2018 and 2019, October and November are 'peak' months where cloud failures are relatively high compared to the rest of the year. During these months cloud failures reach very close to the highest failure count of the year.

\begin{mdframed}[backgroundcolor=gray!20] 
    \textbf{O-1:} The months of October and November had the greatest failure frequency, while December had the lowest number of cloud failures over the years.
\end{mdframed}

In December the cloud failures are the least for both 2018 and 2019. Cloud failures for March have been increasing every year, while in February and May cloud failures have decreased continuously during the three years. The high peak in March 2020 is likely related to Covid-19 the evidence being that this is the first month since October 2018 that cloud failures reach such height (approximately after 1.5 years). This month was the peak lockdown month and most works had been shifted online. The sudden high usage of cloud services during March due to Covid-19 can be a possible explanation for the high peak.

\begin{mdframed}[backgroundcolor=gray!20] 
    \textbf{O-2:} Covid-19 is likely one of the reasons for the peak of cloud failures in March 2020.
\end{mdframed}

An interesting point is that May 2020 had the least number of cloud failures in three years while March 2020 was among the highest cloud failure months. Overall, the pattern of bars shows that when failures are high in a particular month, they do not continue to remain high in upcoming months (high referring to more than 30 failures). The bars always come down in the next month. 

\section{Comparing Cloud Service Providers} 

\begin{figure}[]
    \centering
    \includegraphics[width=0.7\textwidth, keepaspectratio]{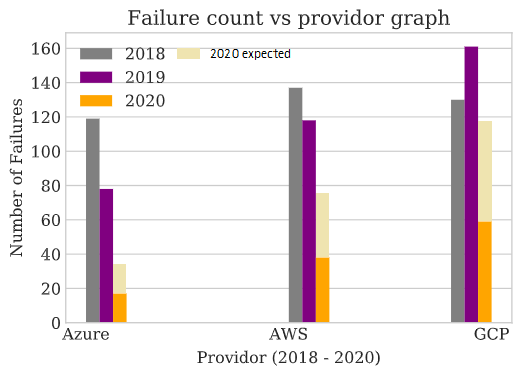}
    \caption{Failure count in cloud service providers.}
    \label{fig:comparing-vendors}
\end{figure}

\begin{mdframed}[backgroundcolor=gray!20] 
    \textbf{O-3:} Microsoft Azure had fewer failures than AWS and GCP.
\end{mdframed}

In 2018 AWS had the most cloud failures, in 2019 GCP had the most cloud failures and in 2020 GCP is expected to have the most failures. Figure \ref{fig:comparing-vendors}  shows that in these three years Azure had fewer failures than AWS and GCP. For both AWS and Azure number of failures decreased from 2018 to 2019. However, the number of failures for GCP increased from 2018 to 2019. Overall, during the 2 and a half-year Azure has fewer failures compared to other vendors.
\newpage

\section{Failure duration and recovery} 

From the plots in Figures \ref{fig:ecdf-2020}, \ref{fig:ecdf-2019} and \ref{fig:ecdf-2018} it can be observed that AWS and GCP recover 70 to 90 percent of failure events in 5 hours while Azure recovers only 50 to 60 percent. In general, during the 2 and a half-year time, AWS has the quickest recovery of cloud failures. GCP recovery speed although behind but is close to AWS. For Azure, the failure event recovery is slower compared to other two vendors. 

\begin{mdframed}[backgroundcolor=gray!20] 
    \textbf{O-4:} AWS had the fastest recovery, followed by GCP. Microsoft Azure recovery was more delayed compared to the other vendors. For all vendors approximately 50\% of cloud failures were recovered in less than 5 hours.
\end{mdframed}

Overall, AWS outages have lasted no longer than 24 hours. GCP had the longest failure occurrence in 2019, spanning up to ten days. In particular, the longest failure occurrence in 2018 was of Azure, which lasted 120 hours. The longest failure occurrence in the first half of 2020 was of GCP, which lasted 29 hours. Eighty percent of cloud outages were recovered within ten hours for all three vendors. In addition, half of the failure events lasted fewer than 5 hours.

\begin{figure}[H]
    \centering
    \includegraphics[width=0.8\textwidth,keepaspectratio]{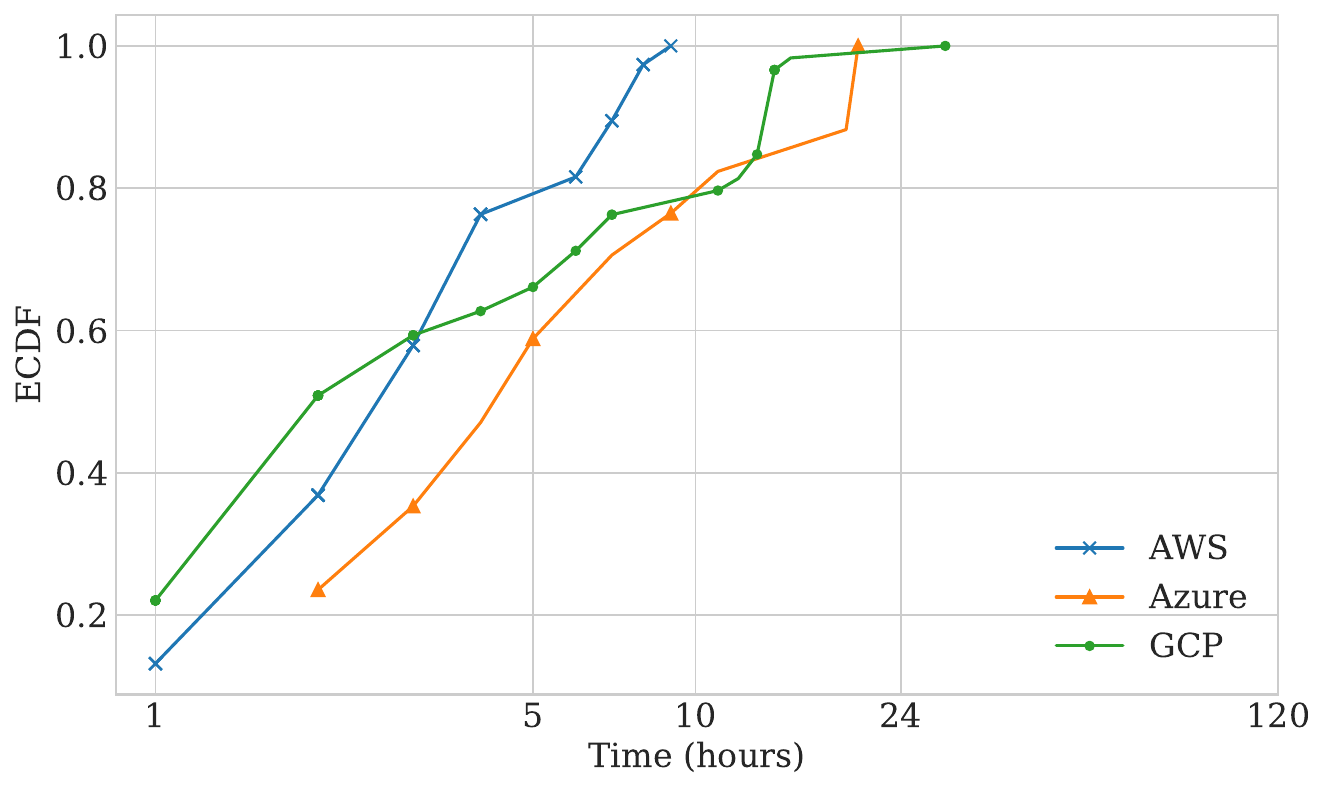}
    \caption{ECDF plot for 2020.}
    \label{fig:ecdf-2020}
\end{figure}%
\begin{figure}[]
    \centering
    \includegraphics[width=0.8\textwidth,keepaspectratio]{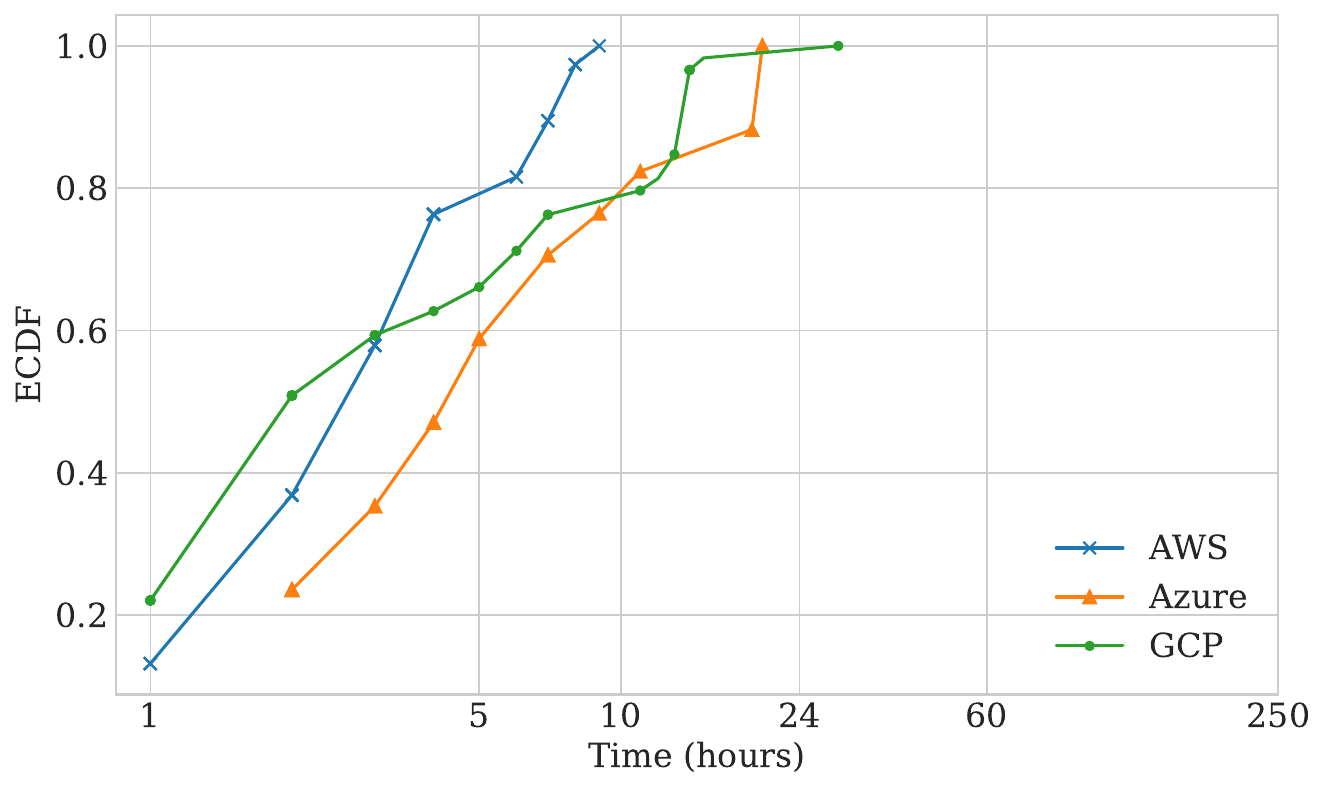}
    \caption{ECDF plot for 2019.}
    \label{fig:ecdf-2019}
\end{figure}

\begin{figure}[]
    \centering
    \includegraphics[width=.8\textwidth,keepaspectratio]{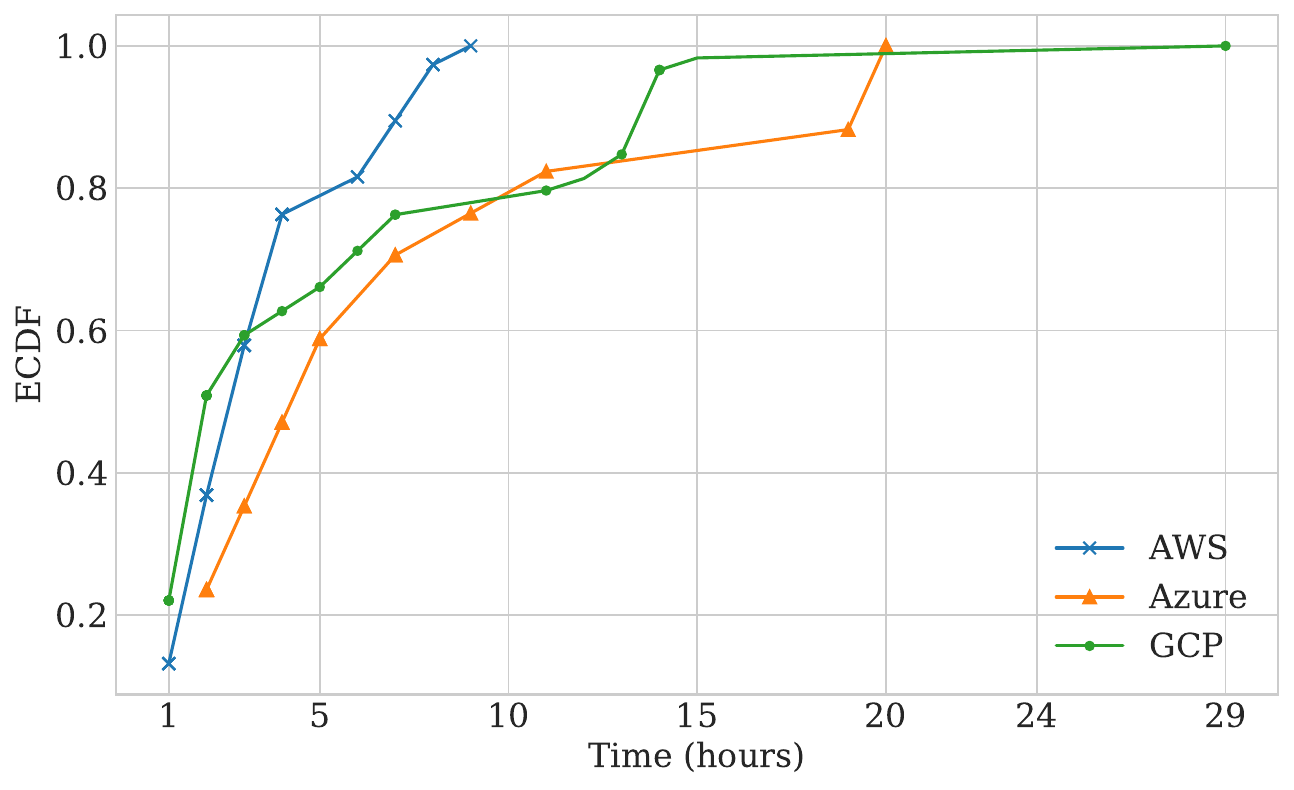}
    \caption{ECDF plot for 2018.}
    \label{fig:ecdf-2018}
\end{figure}

\newpage

\section{Failing Services}

Figures \ref{fig:services-2020}, \ref{fig:services-2019} and \ref{fig:services-2018} shows the name and frequency of the ten services that had most failures in 2020, 2019 and 2018 respectively. A common service failing in all three years is Network service, this is the most failing service throughout the 2.5 years and this service also has the highest number of failures in 2018 and 2019. Similarly Amazon Elastic Compute Cloud service is the second most failing service in all 2.5 years. 

\begin{mdframed}[backgroundcolor=gray!20] 
    \textbf{O-5:} Network service was among the frequently failing service and in total had the most failures over the period. There are many services with continuing failures over the years.
\end{mdframed}

Services that were failing for two consecutive years (2018 and 2019) include Google Cloud Storage service, Google Stackdriver, Google App Engine and Google Compute Engine. In most cases the number of failures for these services have decreased compared to the previous year. Services that have been failing for exactly two consecutive years (2019 and 2020) include Google Cloud Functions and Google Cloud Console.

\begin{figure}[]
    \centering
    \includegraphics[width=\textwidth,keepaspectratio]{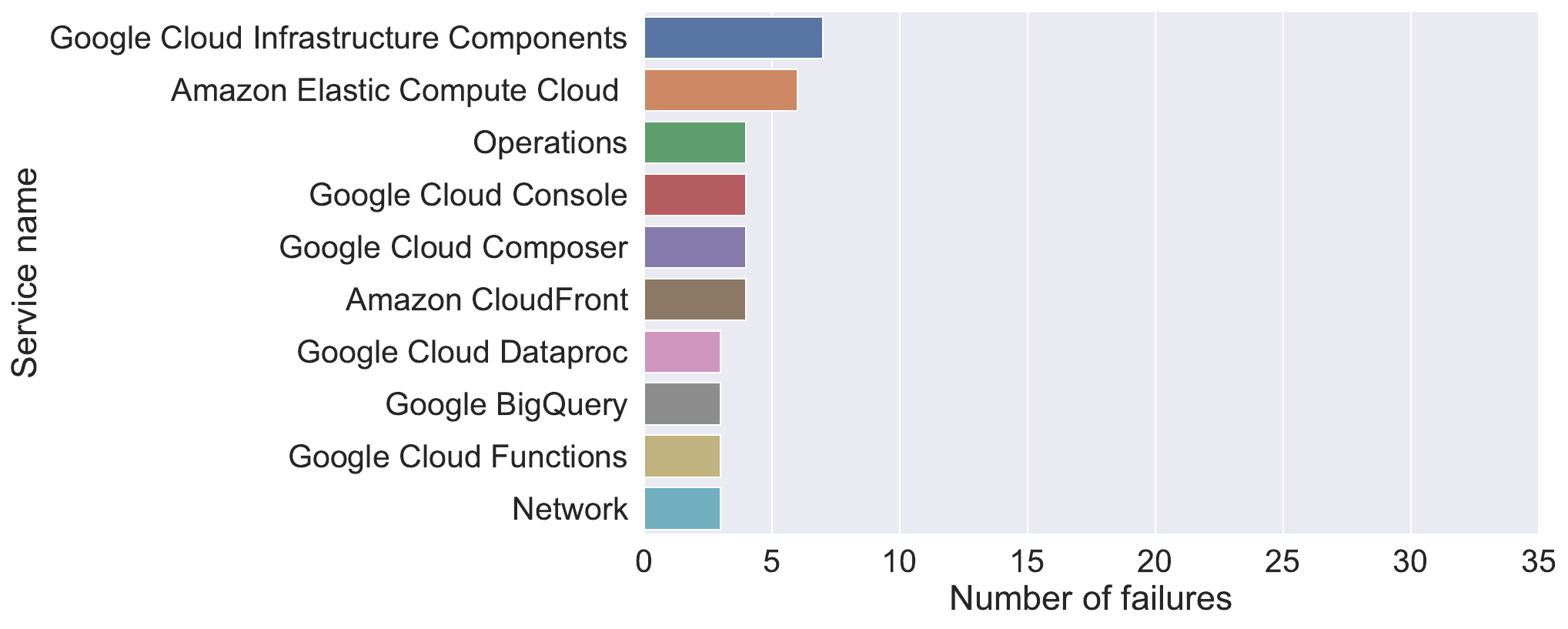}
    \caption{2020 services with most failures.}
    \label{fig:services-2020}
\end{figure}

\begin{figure}[]
    \centering
    \includegraphics[width=\textwidth,keepaspectratio]{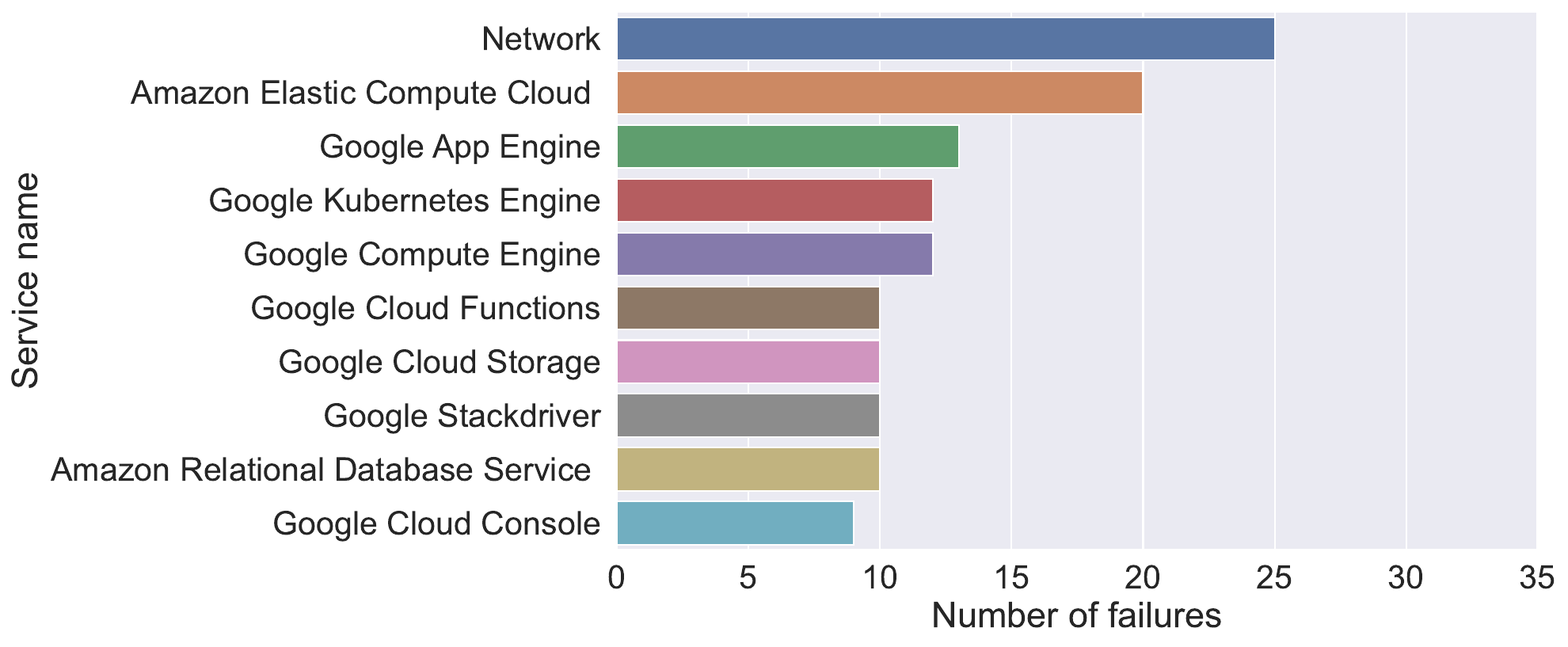}
    \caption{2019 services with most failures.}
    \label{fig:services-2019}
\end{figure}

\begin{figure}[]
    \centering
    \includegraphics[width=\textwidth,keepaspectratio]{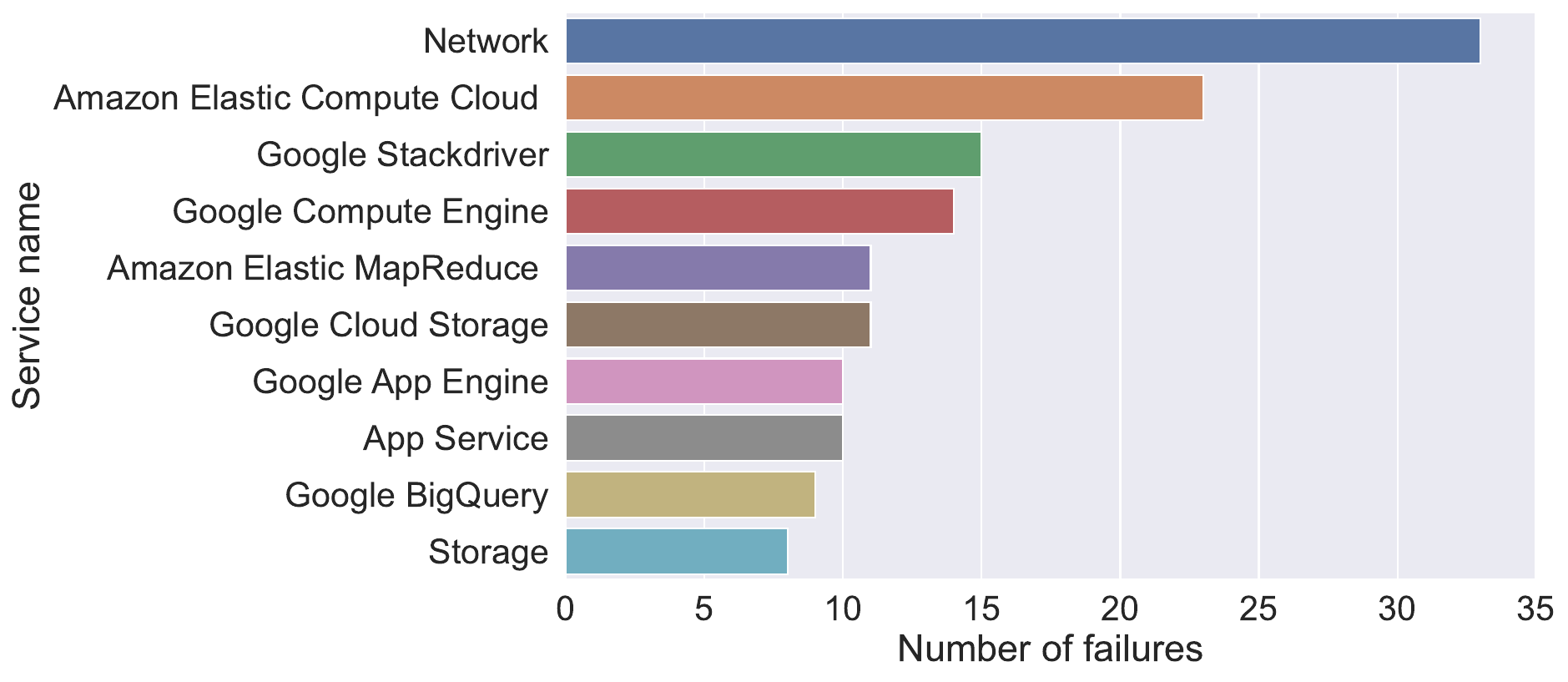}
    \caption{2018 services with most failures.}
    \label{fig:services-2018}
\end{figure}

\newpage

\section{Weeks with Cloud Failures}

Figures \ref{fig:failure-time-2020}, \ref{fig:failure-time-2019} and  \ref{fig:failure-time-2018} shows the failures during the weeks of the year 2020, 2019 and 2018 respectively. The horizontal axis shows the week number while the red dot shows the failure at the day of the week. In 2020, the highest red dot in week 13 can be clearly noticed indicating failure of more than 25 services on a day. In 2019, more than 10 failures on a day in the weeks 30, 39 and week 46. Week 46 has the highest failures in a day that is more than 15 failures in a day. Similarly in 2018, more than 10 failures in a day in weeks 5, 7, 33 and 29. Week 29 had the highest failures. In all three plots, we can notice that it is rare to have a week without failures which highlights the importance of the study. From 2018 to 2019 failures are present but in 2019 failures on a day are not high such as majority is fewer than 5 failures, as a result that is 2019 plot is less scattered vertically.

\begin{mdframed}[backgroundcolor=gray!20] 
    \textbf{O-6:} More than 25 services failed on a day in 2020. Furthermore, a week without cloud failure is rare.
\end{mdframed}

\begin{figure}[]
    \centering
    \includegraphics[width=1.0\textwidth,keepaspectratio]{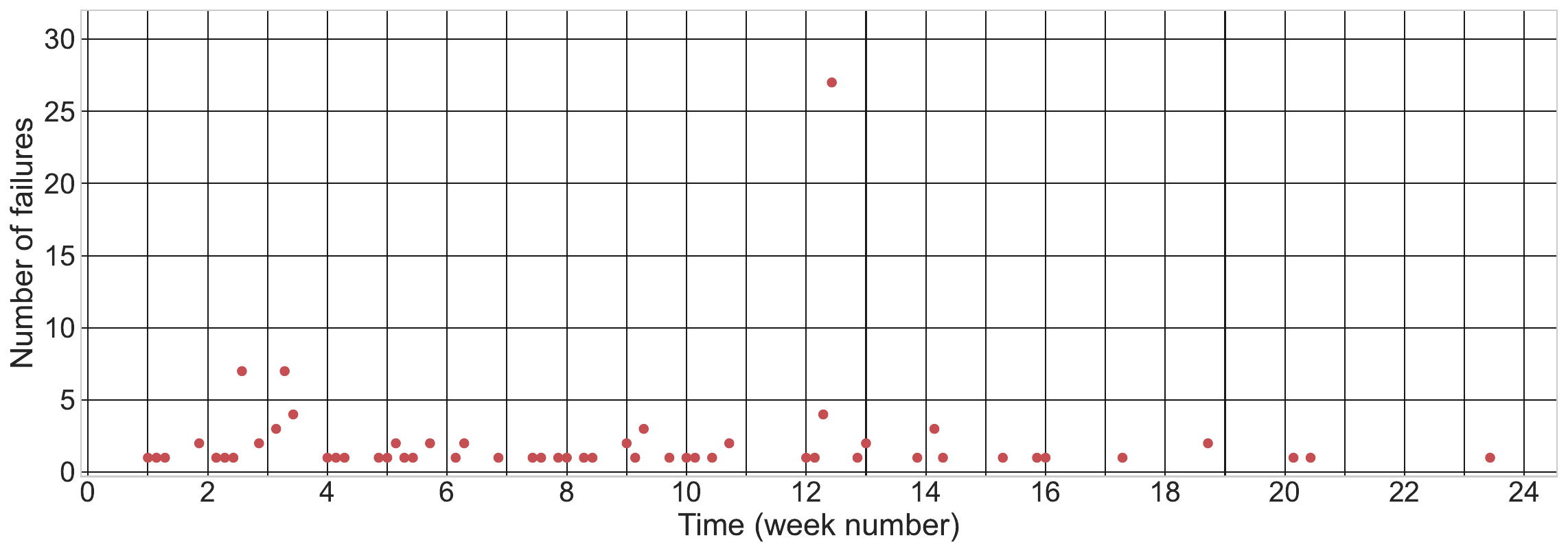}
    \caption{2020 Failures till end of June.}
    \label{fig:failure-time-2020}
\end{figure}

\begin{figure}[]
    \centering
    \includegraphics[width=1.0\textwidth,keepaspectratio]{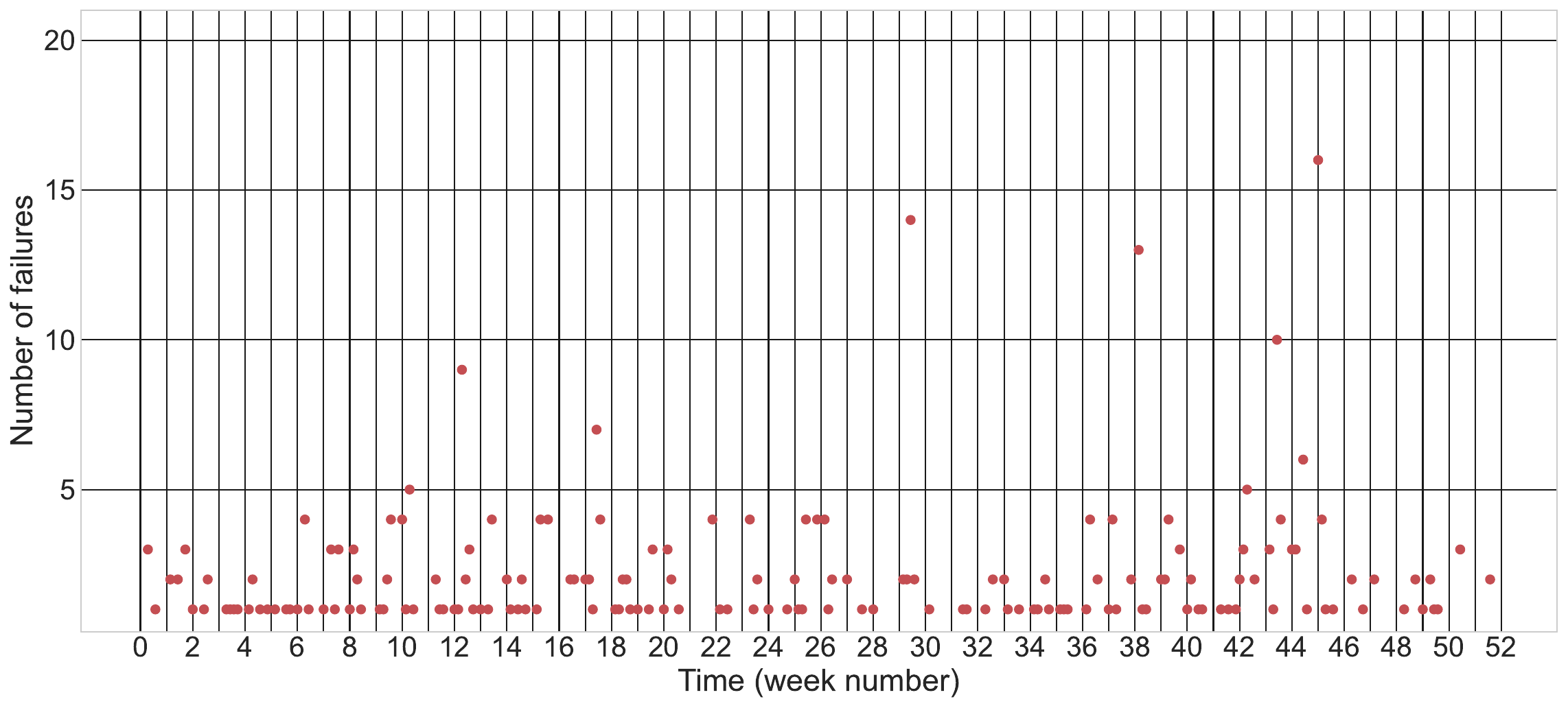}
    \caption{2019 Failures throughout year.}
    \label{fig:failure-time-2019}
\end{figure}

\begin{figure}[]
    \centering
    \includegraphics[width=1.0\textwidth,keepaspectratio]{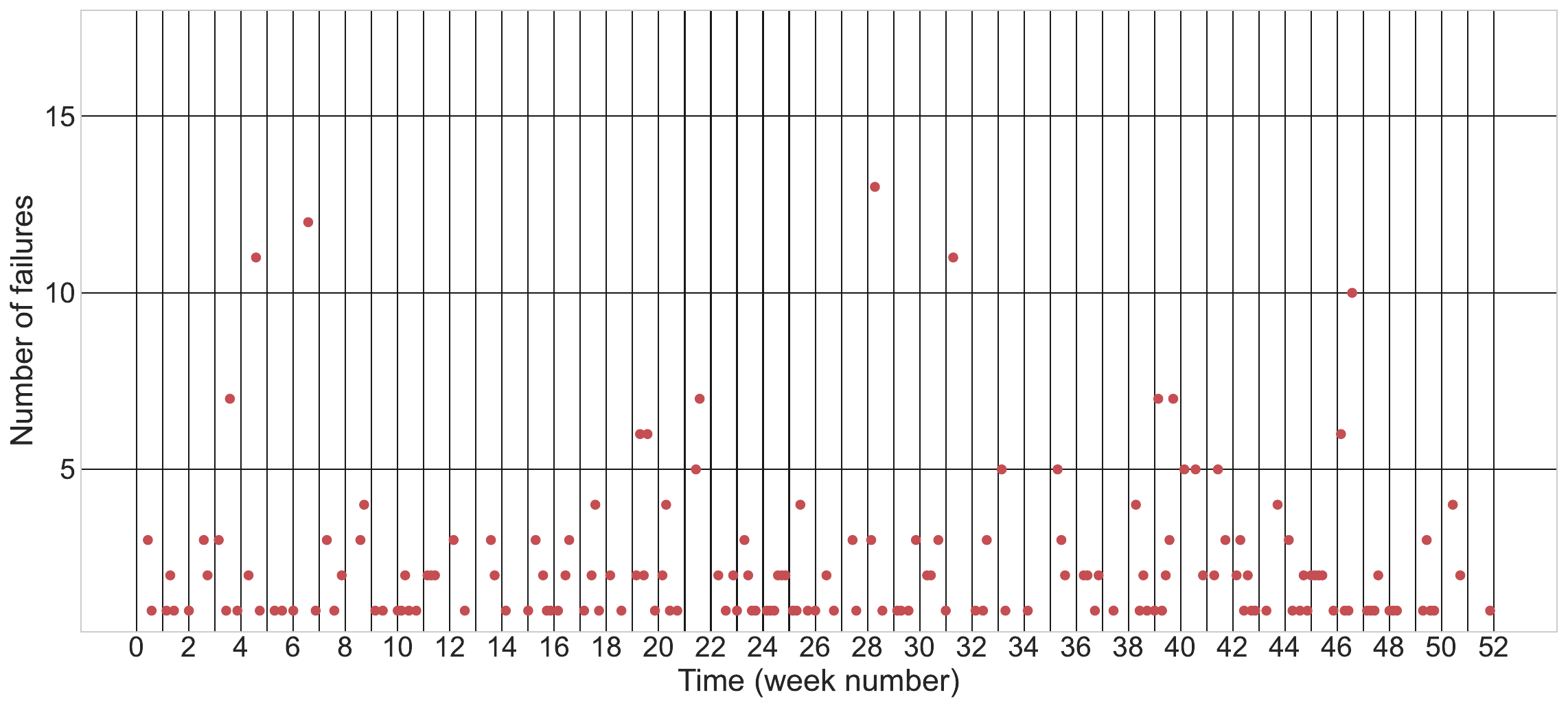}
    \caption{2018 Failures throughout year.}
    \label{fig:failure-time-2018}
\end{figure}

\newpage

\section{Days of Cloud Failures}

Figure \ref{fig:days-2020}, Figure \ref{fig:days-2019} and Figure \ref{fig:days-2018} shows the failures at days of the week in 2020, 2019 and 2018 respectively. In 2018 and 2019 we can observe and it is also expected that more failure occur during working days (Monday to Friday) than on weekend (Saturday and Sunday). In 2020 the pattern is present but may not be clear as only half year data. For both 2018 and 2019 Thursday has the highest number of failures. While in 2020 Friday has the highest number of failures followed by Thursday. Throughout the 2.5 years Sunday has the least number of failures.

\begin{mdframed}[backgroundcolor=gray!20] 
    \textbf{O-7:} Highest cloud failures around Thursday and Friday while lowest on Sunday.
\end{mdframed}

\begin{figure}[]
  \centering
    \includegraphics[width=0.7\textwidth,keepaspectratio]{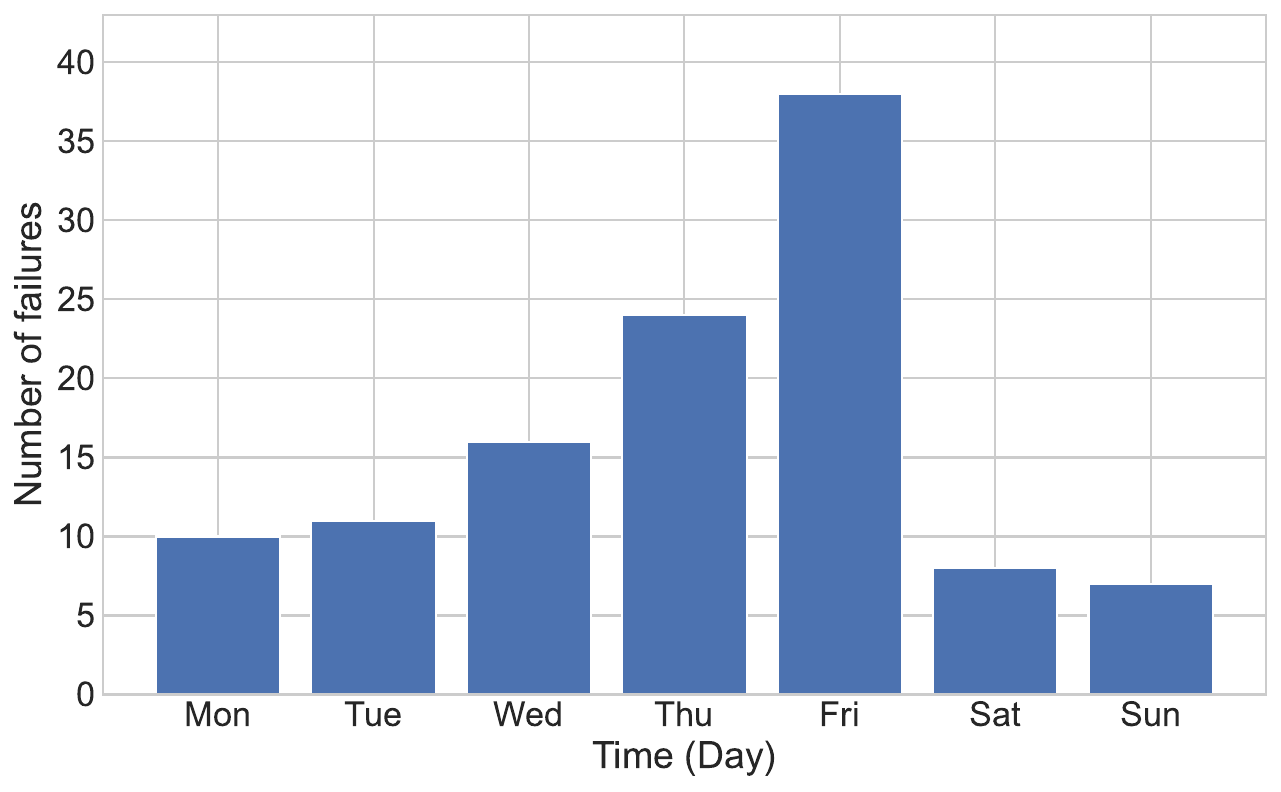}
    \caption{2020 Failures in week days.}
    \label{fig:days-2020}
\end{figure}%
\begin{figure}[]
  \centering
    \includegraphics[width=0.7\textwidth,keepaspectratio]{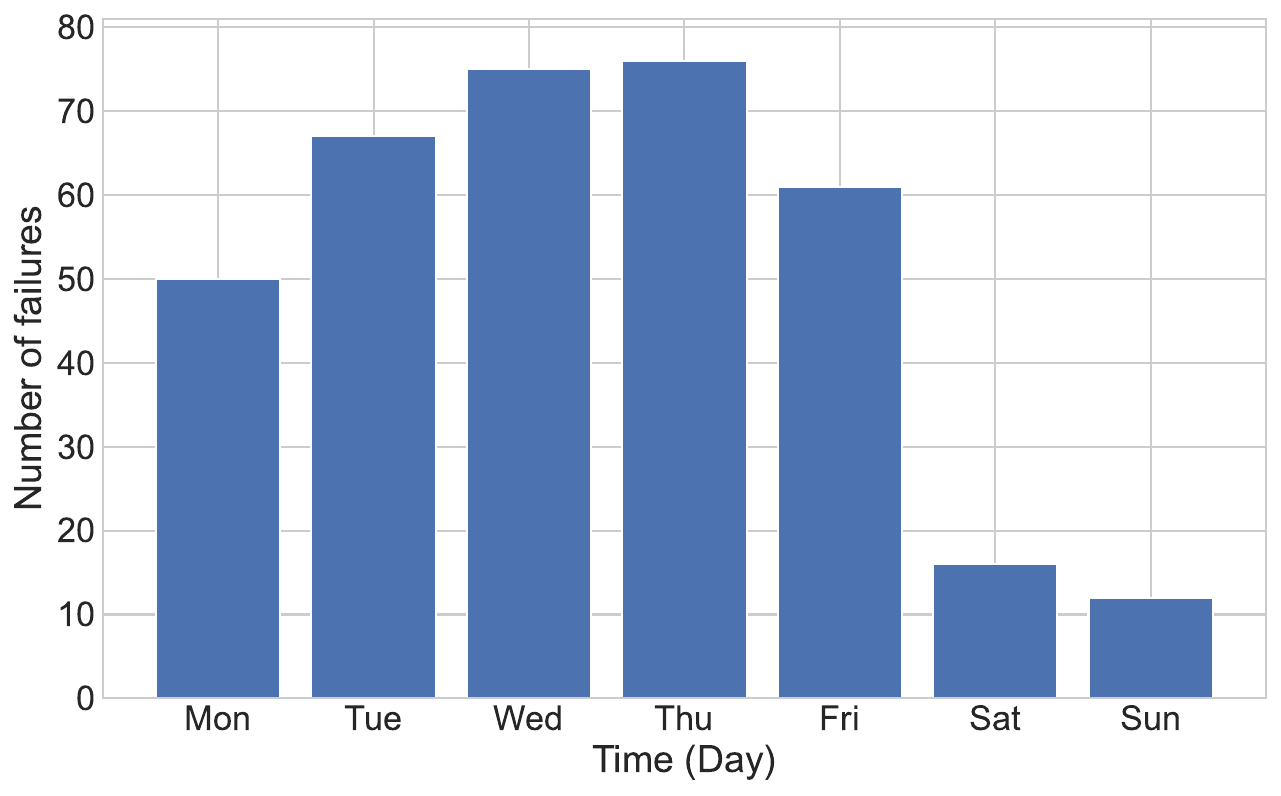}
    \caption{2019 Failures in week days.}
    \label{fig:days-2019}
\end{figure}

\begin{figure}[]
    \centering
    \includegraphics[width=0.7\textwidth,keepaspectratio]{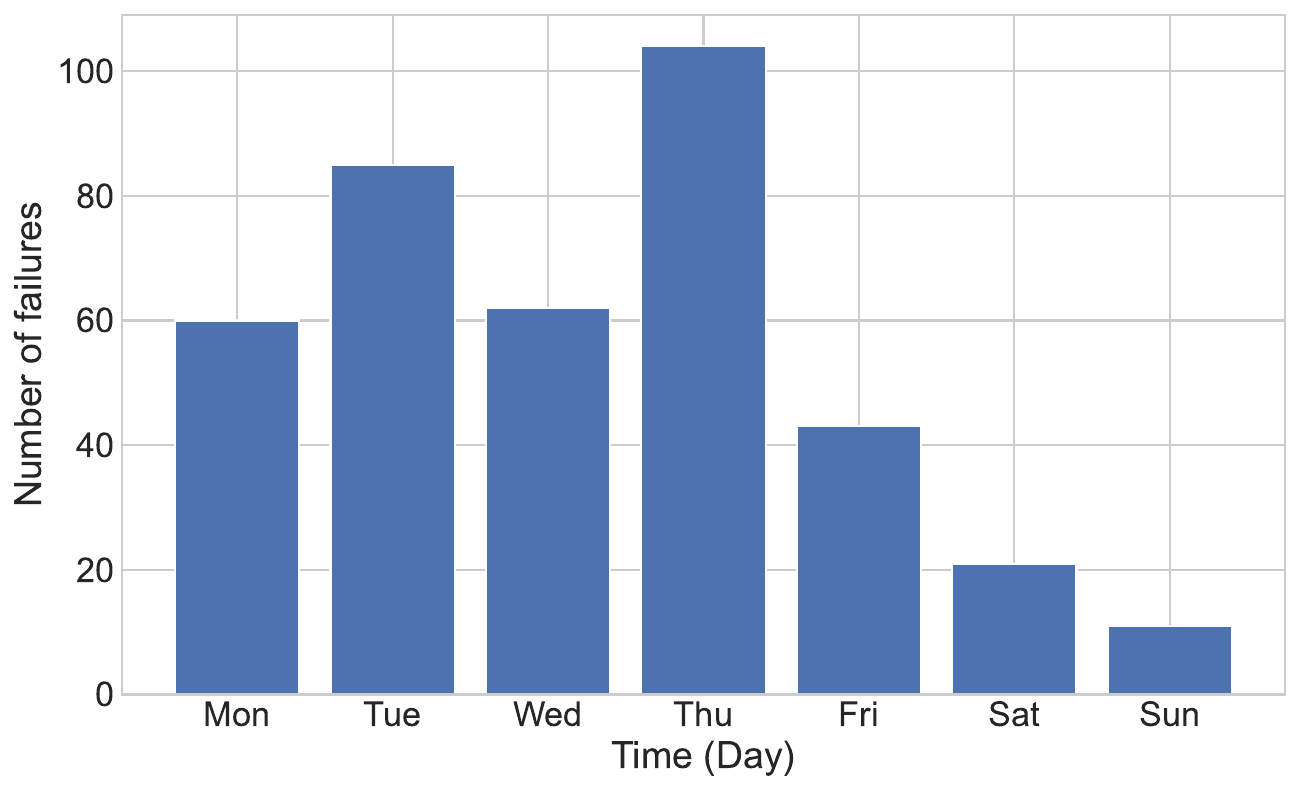}
    \caption{2018 Failures in week days.}
    \label{fig:days-2018}
\end{figure}

\newpage

\section{Location of Cloud failures}

Figures \ref{fig:locations-2020},  \ref{fig:locations-2019} and  \ref{fig:locations-2018} shows the name and frequency of the ten locations that had most failures in 2020, 2019 and 2018 respectively. The location with highest number of failures during the 2.5 years is North Virginia reach 51 failures in 2018 which decrease to 38 in 2019. Another common location throughout 2.5 years is Oregon having above 10 failures during 2018 and 2019 (2020 data is only of half year). Other common locations are East US, Ireland and South US. 

\begin{mdframed}[backgroundcolor=gray!20] 
    \textbf{O-8:} North Virginia and Oregon have the highest and most frequent cloud failures.
\end{mdframed}

The location analysis is limited to the data available that is we do not know locations of all failures. In plots and analysis we also ignore the failure that have multiple origins.

\begin{figure}[H]
    \centering
    \includegraphics[width=.7\textwidth,keepaspectratio]{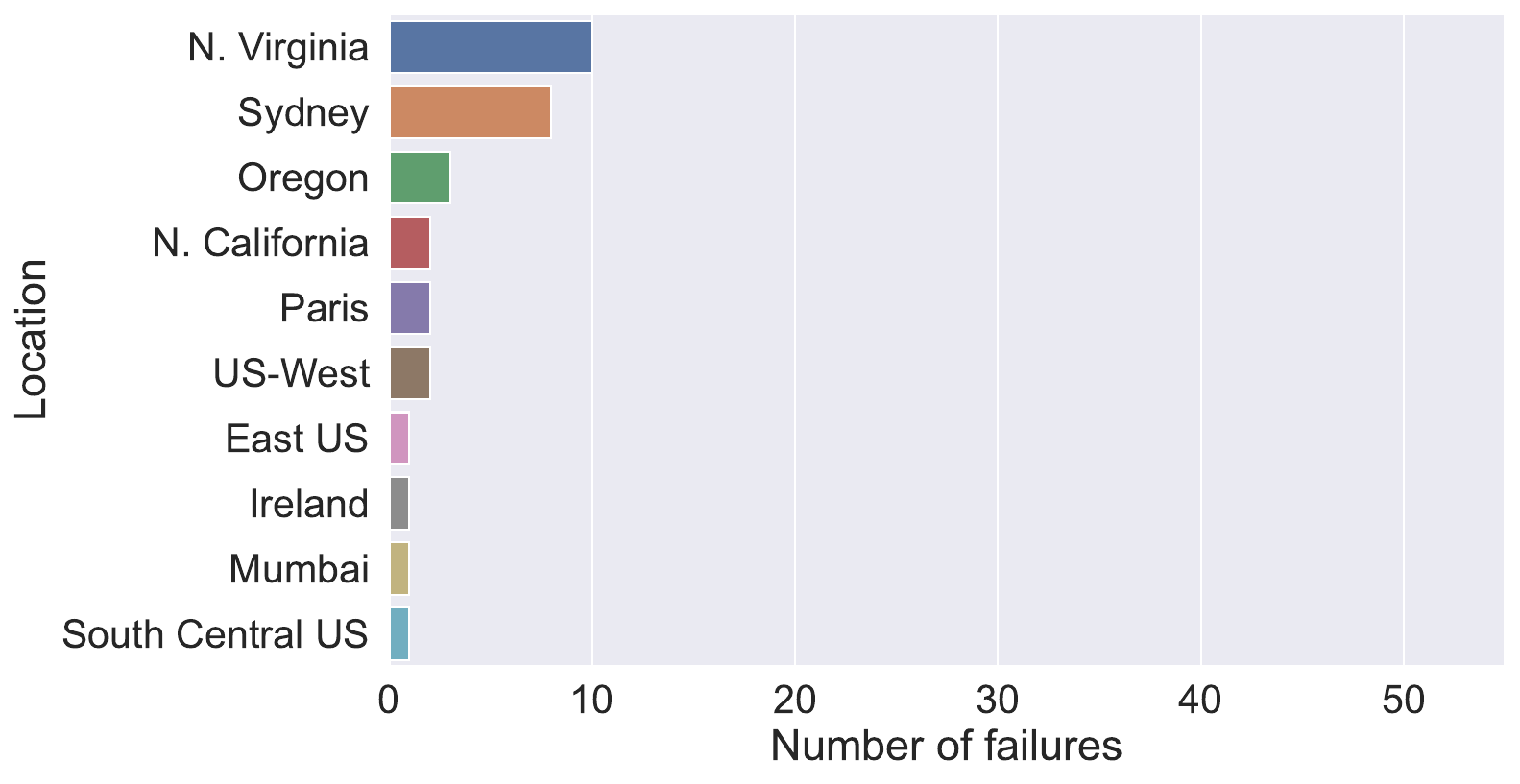}
    \caption{2020 locations with most failures.}
    \label{fig:locations-2020}
\end{figure}%
\begin{figure}[H]
    \centering
    \includegraphics[width=.7\textwidth,keepaspectratio]{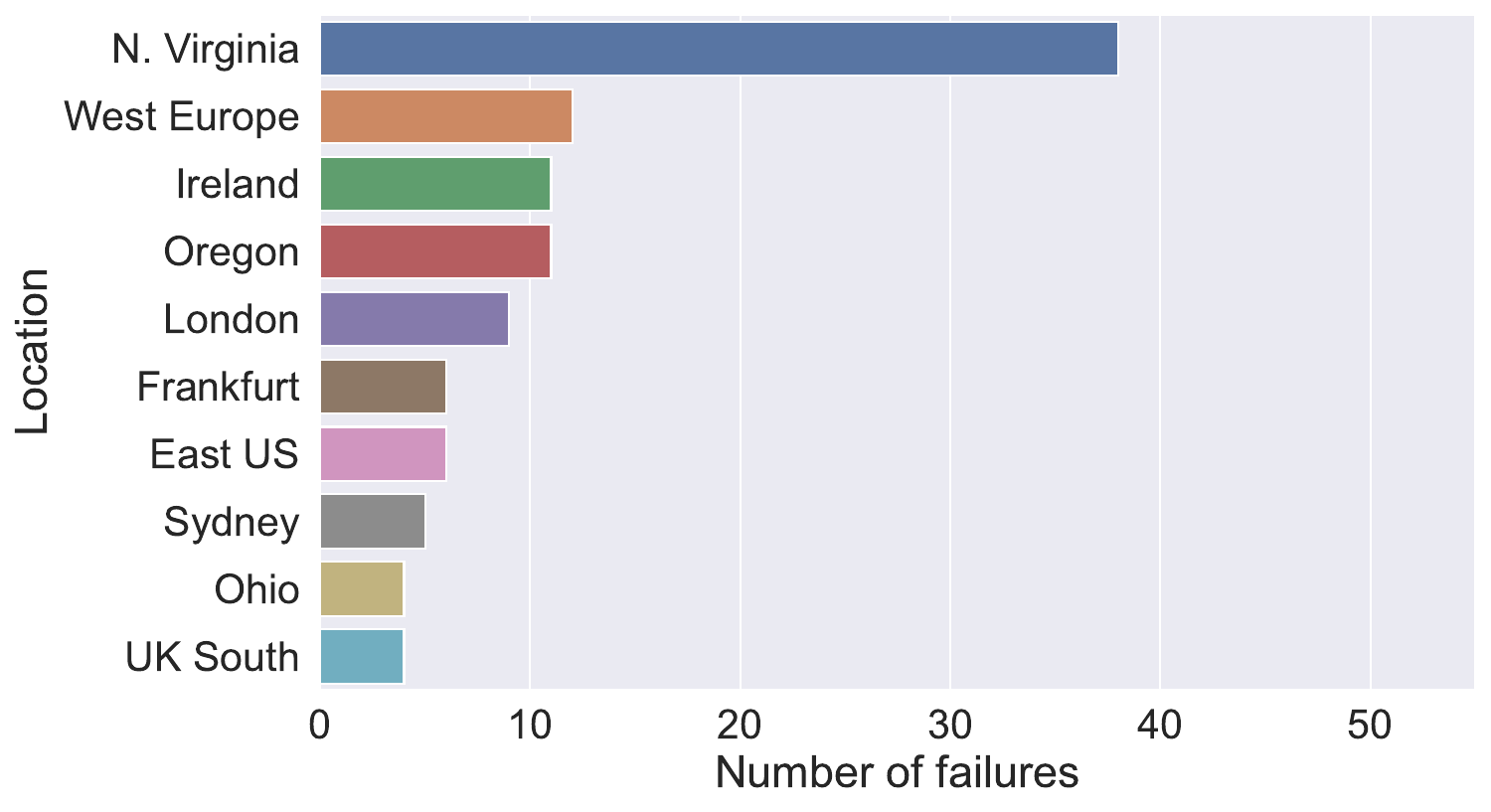}
    \caption{2019 locations with most failures.}
    \label{fig:locations-2019}
\end{figure}

\begin{figure}[H]
    \centering
    \includegraphics[width=.7\textwidth,keepaspectratio]{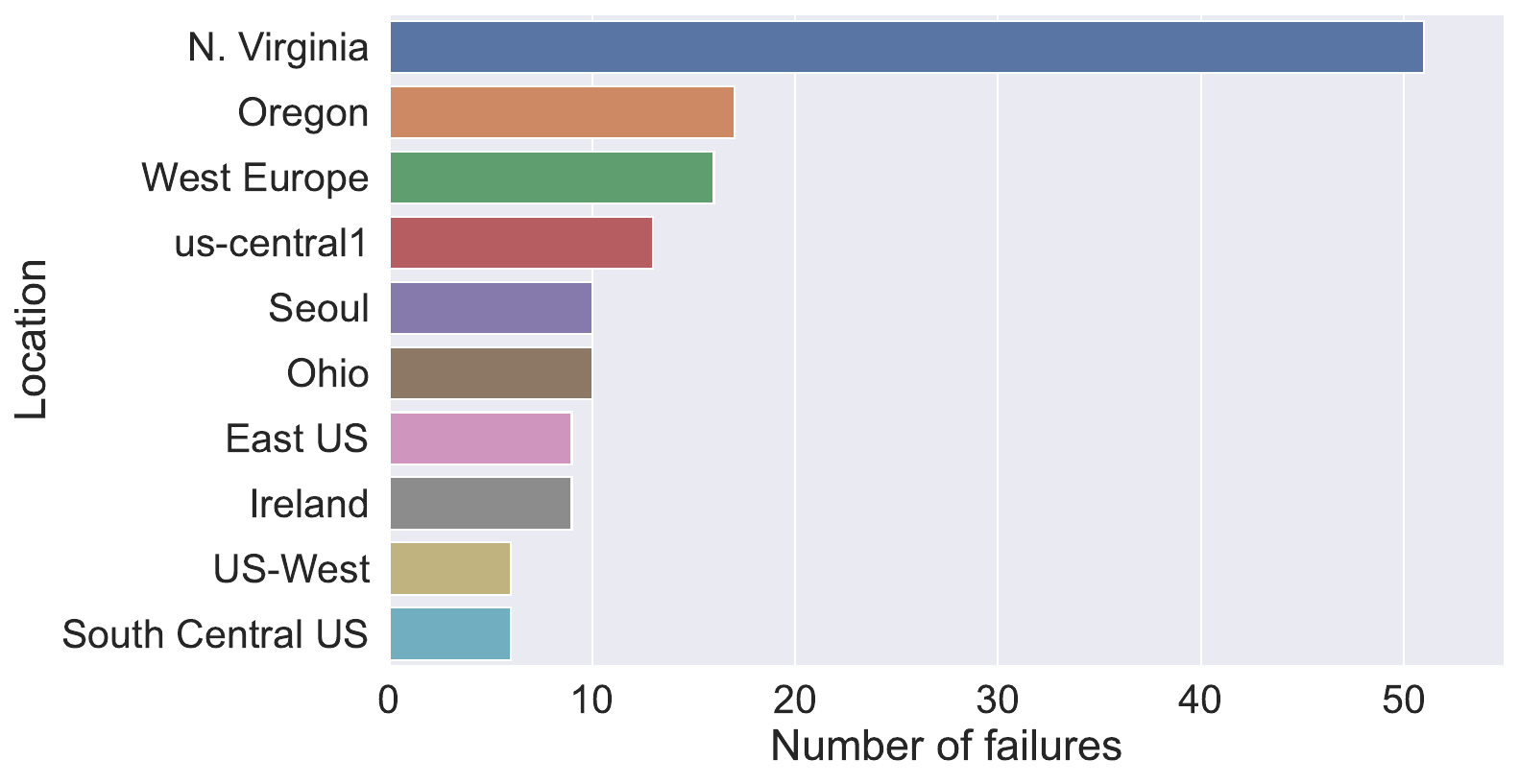}
    \caption{2018 locations with most failures.}
    \label{fig:locations-2018}
\end{figure}

}



{\let\cleardoublepage\relax \chapter{Limitations and Related Work}


\ifpdf
    \graphicspath{{5_evaluation/figures/PNG/}{5_evaluation/figures/PDF/}{5_evaluation/figures/}}
\else
    \graphicspath{{5_evaluation/figures/EPS/}{5_evaluation/figures/}}
\fi

%





This chapter describes the threats to validity in the first section. The second section analyzes related prior work.

\section{Analysis of Limitations} \label{sec:discussion}
The research looks at three different providers; Azure, AWS and GCP. The vendors do not provide all data needed for the study. Missing data is handled in Section \ref{unknown-cells}. Another limitation of the data is that the results depend on the accuracy of the data available. Furthermore, the study only includes data from January 2018 to June 2020 which is two and a half years. When striving for definite conclusions from the data, this quantity of data can be deemed restricted; for example, data from more prior years can improve data comparison.

The study's validity is maintained by using a tool for analysis. The procedure was repeated multiple times for the three files to ensure that the results produced were consistent. Manual inspections were also carried out by examining the program outputs and comparing them to the expected (manual) output. Moreover, the majority of operations are automated, which reduces the risk of human error. Automated validation checks were built in the tool to ensure valid results.

\section{Related Work} \label{sec:related}

Currently, only a limited amount of work has attempted to analyze cloud
failures of big vendors. For example, the authors of \cite{new-outlet-study-2013} analyze outages and incidents reported by companies and news outlets. The current study's data was gathered from official sources \cite{Azuredata, GCPdata, AWSdata} of the vendors. Furthermore, the current study is one of the first because it is based on recent periods. There are studies that use data from newspaper articles to analyse cloud failures. The study gets cloud failure data from official vendor sites. No study exists that examines the failures of the three major cloud service providers, AWS, Azure, and GCP.

}




{\let\cleardoublepage\relax \chapter{Conclusion} \label{sec:conclusion}


\ifpdf
    \graphicspath{{8_conclusion/figures/PNG/}{8_conclusion/figures/PDF/}{8_conclusion/figures/}}
\else
    \graphicspath{{8_conclusion/figures/EPS/}{8_conclusion/figures/}}
\fi

%



This chapter summarizes the work and contributions of the study. Cloud services are beneficial to people all around the world. The cloud is viewed as a solution to a variety of issues. For example, the cloud allows people to collaborate and communicate with one another, particularly when they are in different countries. There is, however, a long list of cloud failures that could have a detrimental impact on billions of cloud users. Understanding cloud failures is essential. 

We created a tool to analyze and understand cloud failures. Initially the tool was made to process raw cloud failure data. The tool was then extended to output statistics and visuals from the data. These are then analyzed. Comparison of failures prior to and during Covid-19 is studied. By studying cloud failures, many of cloud failures can be prevented along with the loss caused by cloud failures. This study covered the period 2018 till 2020-June and provided analyses of cloud failures in big cloud providing companies; AWS, Microsoft Azure and GCP. Furthermore, this study provides insight for reasons of cloud failures. For further study, the tool can be extended to be used for analysis of future cloud failures. 

}



{\let\cleardoublepage\relax \chapter{Self-Reflection}}


\ifpdf
    \graphicspath{{7_lessonslearned/figures/PNG/}{7_lessonslearned/figures/PDF/}{7_lessonslearned/figures/}}
\else
    \graphicspath{{7_lessonslearned/figures/EPS/}{7_lessonslearned/figures/}}
\fi

%

\section{Self-Reflection} \label{app:selfreflection}

Through this project I learned about growing demand and importance of cloud services. It was interesting to learn about cloud failures including why, where, how they occur and their possible solutions. During the research I interacted with many professional members of the team that guided me through out the study. I was greatly inspired by their words that introduced my to methodologies that I had not known. The research showed me a variety the techniques to analyse data. This study boosted my skills as a python programmer as I gained experience by creating a software for the study. I learned programming skills used for data analysis, applying statistical methods, plotting graphs using code, various ways of representing data and different types of graphs such as the ECDF plot. The most attractive part of the study was the results of analysis. \newline

During the research two-third of the time was spent on data cleaning and visualization. Among which more time was spent by cleaning data (RQ1), than on data transformation (RQ2). The data provided was raw data. Converting the raw data in use-able data was done using a software created during the study. Furthermore cleaning data involved many operations, see Section \ref{cleaning_data} The leftover one-third time was spent on analysis and documentation.







\bibliographystyle{Latex/Classes/PhDbiblio-url2} 
\renewcommand{\bibname}{References} 

{\let\cleardoublepage\relax }










\end{document}